\documentclass[aps,prl,twocolumn,superscriptaddress,groupedaddress,citeautoscript,preprintnumbers,floatfix]{revtex4-2}
\usepackage[utf8]{inputenc}
\usepackage[T1]{fontenc}
\usepackage{float} 
\usepackage{color}  
\usepackage{graphicx}
\usepackage{amsmath}
\usepackage{amssymb}
\usepackage{bbm}
\usepackage{indentfirst}
\usepackage{dcolumn}
\usepackage{soul}
\usepackage{mathrsfs}
\usepackage{xcolor}
\usepackage{ulem}

\usepackage{relsize}
\usepackage{amsmath}
\usepackage{graphicx}
\usepackage{float}

\usepackage{color}
\makeatletter
\usepackage{url}
\usepackage{hyperref}
\hypersetup{
    colorlinks=true,
    urlcolor=blue,
    linkcolor=blue,
    citecolor=blue}

\usepackage{xcolor}
\usepackage{soul}
\usepackage{soulpos}

\usepackage{dcolumn}
\usepackage{bm}
\usepackage{subfigure}
\usepackage{helvet}
 \usepackage{color}
\graphicspath{{figures/}}

\begin{document}

\title{Atomistic spin model of single pulse toggle switching in Mn$_2$Ru$_x$Ga Heusler alloys}

 \author{F. Jakobs}
\affiliation{Dahlem Center for Complex Quantum Systems and Fachbereich Physik, Freie Universit\"{a}t Berlin, 14195 Berlin, Germany}
\author{U. Atxitia}
\email{unai.atxitia@fu-berlin.de.}
\affiliation{Dahlem Center for Complex Quantum Systems and Fachbereich Physik, Freie Universit\"{a}t Berlin, 14195 Berlin, Germany}

\begin{abstract}
Single femtosecond pulse toggle switching of ferrimagnetic alloys is an essential building block for ultrafast spintronics. Very different element-specific demagnetization dynamics is believed to be a hard limit for switching in ferrimagnets. This suggests that ferrimagnets composed of two ions of different nature, such as rare earth transition metal alloys, are necessary for switching. However, experimental observation of toggle switching in Mn$_2$Ru$_x$Ga Heusler alloys, has contested this limit since Mn ions are of the same nature. To shed some light into this question, we present an atomistic spin model for the simulation of single pulse toggle switching of Mn$_2$Ru$_x$Ga. The magnetic parameters entering in our model are extracted from previous experimental observations. We show that our model is able to quantitatively reproduce measured magnetization dynamics of single pulse toggle switching. We demonstrate that differently to previous understanding toggle switching in Mn$_2$Ru$_x$Ga is possible even when both Mn sublattices demagnetization at very similar rate.  
\end{abstract}

\maketitle

Single pulse femtosecond toggle switching in ferrimagnets has attracted a lot of attention as a promising solution for low energy, faster memory applications.~\cite{OstlerNatComm2012,Kimel2019,El-Ghazaly2020,Barker2021}. It has already been demonstrated in micro and nanostructures \cite{LeGuyaderNatComm2015, El-Ghazaly2019}, to switch magnetic tunnel junctions \cite{Chen2017}, as passive component to induce switching in ferromagnets \cite{Gorchon2017} and even using picosecond electric pulses \cite{YangSciAdv2016}.
Toggle switching has been mostly found in one class of material,  systems composed of transition metals rare-earth, e.g. GdFeCo \cite{OstlerNatComm2012} and TbFeCo \cite{LiuNanoLetters2015} alloys, and Gd/Co\cite{Lalieu2017} and Tb/Co stacks \cite{Aviles-Felix2020}. Integrating all optical switching with spintronics using this class of material has been proposed\cite{Lalieu2019}, however, Mn$_2$Ru$_x$Ga alloys, the second class of material showing toggle switching, is better suited \cite{banerjee2019}.
While for GdFeCo  intense research has provided a large amount of experimental data permitting the validation of a number of theoretical models, for Mn$_2$Ru$_x$Ga, there only exist a few experimental works showing ultrafast magnetization dynamics and switching \cite{Bonfiglio2019,banerjee2019,Bonfiglio2021,Banerjee2021}. 

\begin{center}
\begin{figure}[t]
\includegraphics[width=0.9\columnwidth]{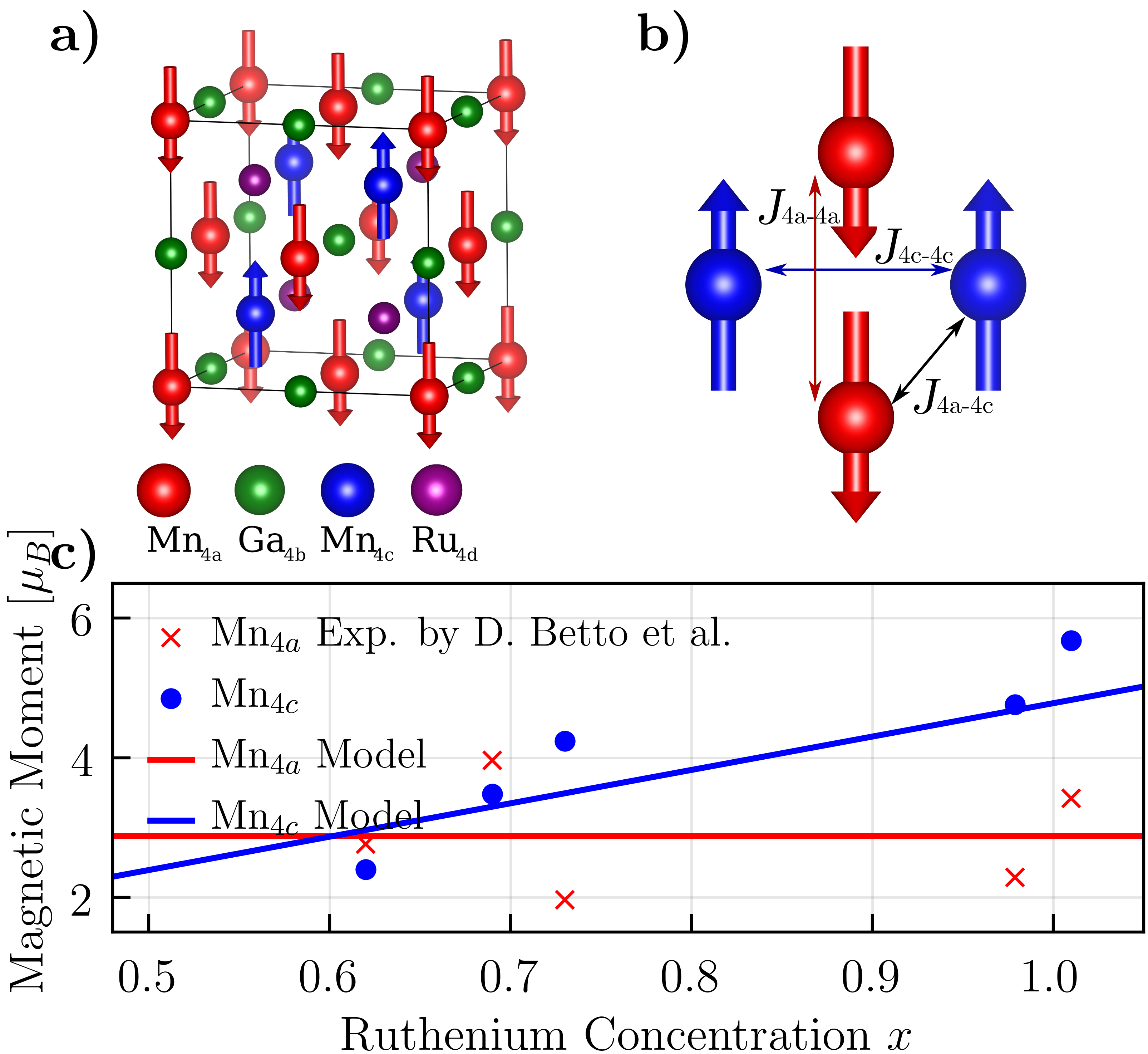}
\caption{a) Crystal structure of Mn$_2$Ru$_x$Ga with the two magnetic Mn-sublattices represented as blue and red arrows. b) Schematic of the exchange constants $J_{4a-4a}$, $J_{4a-4c}$ and $J_{4a-4c}$. c) Mn-specific atomic magnetic moments, symbols correspond to experimental data~\cite{Betto2015} and lines to the values used in the atomistic spin model.}
\label{fig:MaterialProperties}
\end{figure}
\end{center}

The Mn$_2$Ru$_x$Ga Heusler alloy crystallises in the cubic space group F$\overline{4}3m$ with the magnetic Mn atoms on the $4a$ and $4c$ sites (Fig. \ref{fig:MaterialProperties}a). 
Spins at $4a$ and $4c$ sites are coupled antiferromagnetically, whereas the respective Mn $4a$- and $4c$ sublattices are coupled ferromagnetically \cite{Fowley2018}. 
The Ga atoms appear at the $4b$ sites and Ru atoms at the $4d$ sites of the lattice, and their magnetic contribution can be neglected. 
Ultrafast magnetization dynamics of the magneto-optically active Mn $4c$ sublattice was first measured by Bonfiglio et al. \cite{Bonfiglio2019}.
In a subsequent work, they concluded that Mn $4c$ sublattice shows an ultrafast demagnetization ($\sim 100$s fs) followed by either a secular equilibrium or by a fast remagnetization ($\sim 1$ ps). 
By using a phenomenological model based on the so-called four temperature -- electron, phonon, and spin temperatures of Mn $4a$ and $4c$ -- these dynamics were interpreted as a signature of strong exchange-driven relaxation \cite{Bonfiglio2021}. 
Spin temperature models however are unable to describe angular momentum transfer and switching.
Experiments carried out by Davies et al. demonstrated that switching is possible in Mn$_2$Ru$_x$Ga, but only when the initial temperature ($T_0$) lies below the magnetization compensation temperature ($T_{\rm{M}}$)  \cite{Davies2020}.
Based upon a phenomenological model for the magnetization dynamics of two-sublattice magnets \cite{MentinkPRL2012}, the authors argued that in order to link static thermodynamic properties (equilibrium $T_{\rm{M}}$) to highly non-equilibrium dynamics (switching), both sublattices should demagnetize at the same rate, conserving the total angular momentum during the whole process. This picture explains their observation of the switching onset for a wide range of system parameters, including switching with picosecond-long pulses. This interpretation collides with the phenomenology proposed by Bonfiglio et al.  \cite{Bonfiglio2021}, which states that both sublattices only demagnetize at the same rate after $\sim$ 1.5 ps. 
 Still, since data on the magnetization dynamics of the switching process was missing, the question remained open. An apparent step forward along this direction was made by Banerjee et al. \cite{banerjee2019} by measuring the dynamics of the Mn $4c$ sublattice when switching occurs.  They were unable to observe an explicit switching of the sign of the magneto-optical signal, however the dynamics of the Mn $4c$ sublattice behaved like those observed by Bonfiglio\cite{Bonfiglio2021}. 
These observations differ strongly with the well-known element specific signal switching measured in GdFeCo \cite{Radu2012}. Recently, however, Banerjee and co-workers have clearly observed the dynamics of magnetization switching  of the Mn $4c$ sublattice \cite{Banerjee2021}, in clear disagreement to their own previous observations~\cite{banerjee2019}.  
This disagreement could be related to the different strength of the resetting magnetic field, stronger in the latter.
Similarly to the previous works\cite{banerjee2019,Davies2020,Bonfiglio2021}, the understanding of the physics behind the switching process rested on phenomenological arguments. 
An attempt to describe switching in Mn$_2$Ru$_x$Ga using a first-principles model exists \cite{Zhang2020}. Although this model provides useful insights on the potential origin of  switching, due to its simplicity it is unable to describe the temperature dependence of the switching condition ($T_0<T_M$) observed by Davies et al. \cite{Davies2020}  and Banerjee et al. \cite{banerjee2019}. A quantitative model able to account for thermodynamic aspects of the magnetization switching dynamics in Mn$_2$Ru$_x$Ga is necessary but so far missing.

\begin{center}
\begin{figure}[t]
\includegraphics[width=1\columnwidth]{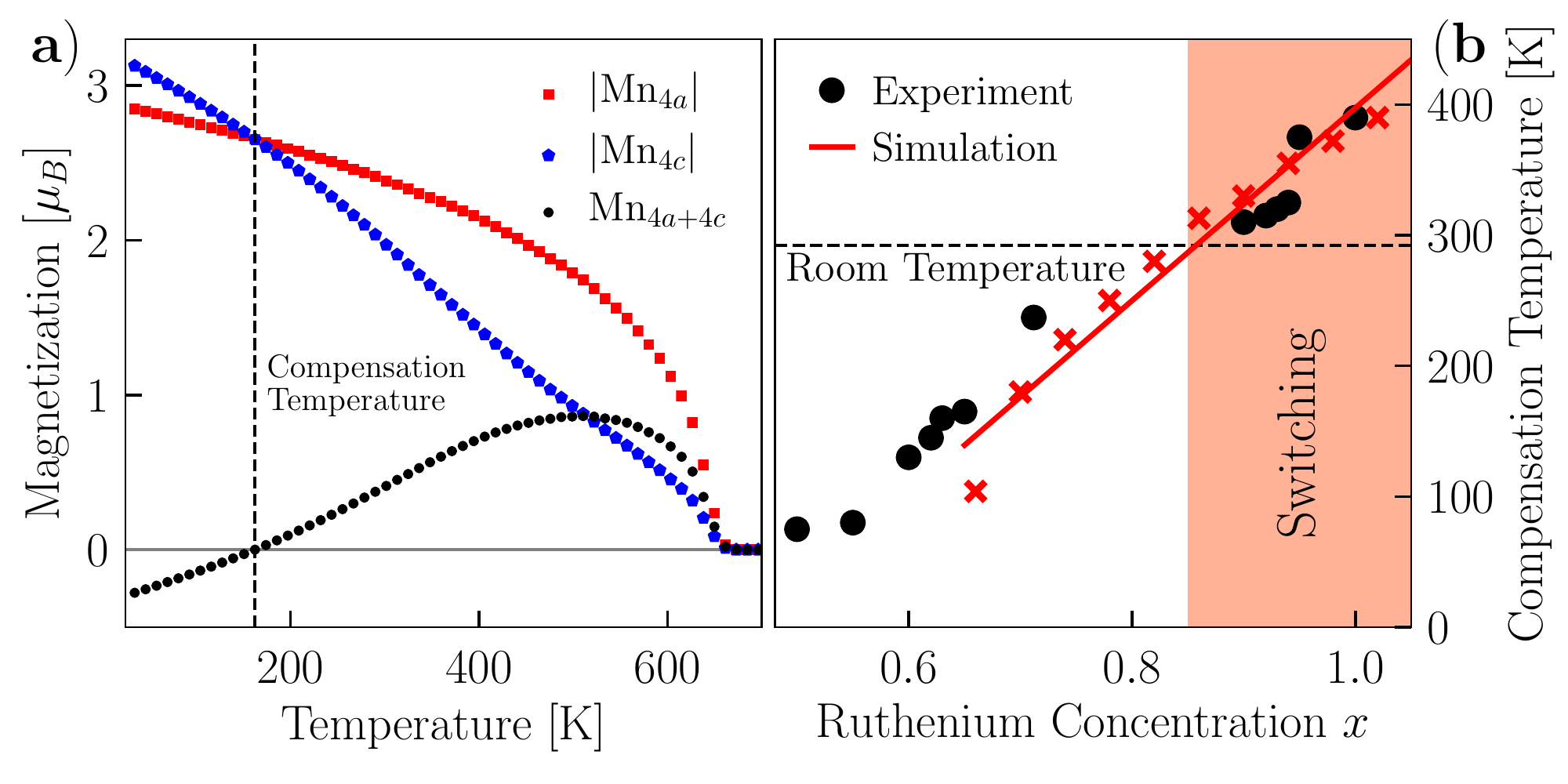}
\caption{ Magnetic  properties of Mn$_2$Ru$_x$Ga alloys gained from atomistic spin model simulations. a) The equilibrium magnetization for a Mn$_2$Ru$_{0.68}$Ga alloy as function of temperature of the Mn$_{4a}$ sublattice in red, the Mn$_{4c}$ sublattice  and the total magnetization $\text{Mn}_{4a+4c} = |\text{Mn}_{4a}| - |\text{Mn}_{4c}$|. b) displays the magnetization compensation temperature $T_M$ as function of Ru-concentration, $x$.  Black dots correspond to experimental measurements ~\cite{banerjee2019} and red crosses from our model simulations. The red line is a linear fit of the simulated results and is guidance to the eye.}
\label{fig:eqProperties}
\end{figure}
\end{center}
In this work we address this issue by presenting an atomistic spin model to describe ultrafast magnetization and switching in Mn$_2$Ru$_x$Ga alloys. Atomistic spin models have demonstrated themselves to be able to describe the dynamics of the magnetization switching in transition metal rare-earth alloys\cite{OstlerNatComm2012,Jakobs2021} and multilayers \cite{Xu2016,Gerlach2017}. 
Atomistic spin models are based on a semiclassical spin Hamiltonian, which is defined by the exchange and anisotropy constants as well as the atomic magnetic moments (see Supplemental Material for details). 
 In a minimal model for Mn$_2$Ru$_x$Ga, one needs to determine at least three exchange constants, $J_{4a-4a}$,$J_{4c-4c}$ and $J_{4a-4c}$ (see Fig.~\ref{fig:MaterialProperties}~b)). 
 Within a Heisenberg spin model, the values of the exchange constants determine the critical temperature $T_c$, and in ferrimagnets, besides $T_c$, the magnetization compensation temperature, $T_M$, if it exists.
The reported $T_c$ range from $T_c=625$ K~\cite{Fowley2018} ($x=0.68$), $T_c=450$ K ($x>0.5$)~\cite{Kurt2014}, $T_c=550$ K ($x=0.7$)~\cite{Bonfiglio2021}. We fix $T_c$ to about 630~K  which is at the upper end of reported critical temperatures.
Here, we use values of the exchange constants already determined experimentally~\cite{Fowley2018}, but slightly re-scaled such that our model reproduces the experimental values of both $T_c$ and $T_M$ (Fig. \ref{fig:eqProperties}).
The anisotropy constant in Mn$_2$Ru$_x$Ga has been reported to be site-specific, with $d_{z,4c} = 1.1664 \cdot 10^{-23}$ J and $d_{z,4a} = 0$ \cite{Fowley2018}, which we use in our model. We note that the role of the anisotropy in the ultrafast magnetization dynamics is minimal since $d_z/J \ll 1$ and it is included to fix the average magnetization towards the $z$-axis. 
Any choice of relatively small values of the anisotropy constant  would yield the same simulation results.

Lastly values for the site-specific Mn atomic magnetic moments are needed.
Based upon experimental data, we assume that the atomic moment of the Mn $4a$ sublattice stays constant at $\mu_{s,4a}= 2.88 \mu_{\rm{B}}$ for the range of Ru-concentration studied here~\cite{Betto2015} (Fig. \ref{fig:MaterialProperties}c)). Differently, the Mn $4c$ sites have been shown to have a stronger dependence on the Ru-concentration~\cite{Betto2015}(Fig. \ref{fig:MaterialProperties}c). 
Here we use values of $\mu_{s,4c} = 4.71\mu_B \cdot x$ (Fig. \ref{fig:MaterialProperties}c)) that are both close to the experimentally measured values and provide good results for $T_M$ (see Fig. \ref{fig:eqProperties}). These values not only compare well to those found in experiments\cite{Betto2015,Yang2015} but also agree with the general trend found in first-principles calculations~\cite{Zhang2020,Galanakis2014}. 
Simulations of our model are able to reproduce well the observation of an increasing $T_M$ for an increasing Ru-concentration. For $x<0.66$, our model starts to deviate from the experiments~\cite{banerjee2019} (Fig.~\ref{fig:eqProperties} b)).

The shared wisdom of single femtosecond pulse toggle switching in ferrimagnets is based on experimental observations in only one class of material, transition metal rare-earth compounds. Despite their structural or morphological differences, in those materials the conditions for switching are based in the same working principles, the rare-earth spin sublattice response to the laser pulse is slower than that of the transition metal. The physical reason behind this difference relies in the core character of the highly localized 4$f$ electron spins in Gd, plus the absence of orbital magnetic moment ($L=0$) \cite{FrietschNatComm2015}. The laser only excites them indirectly, by their coupling to the 5d6s itinerant electrons.
The transition metal spins respond quickly to changes in the electronic structure band due to the quick temperature rise.

Based on this it has been argued for long that toggle switching is only possible for two sublattice compounds with components that demagnetize at different enough rates.
Since data on site-specific dynamics is unavailable, it is unclear whether or not this criterion holds in Mn$_2$Ru$_x$Ga.
It is tempting to draw similarities to GdFeCo to explain switching in  Mn$_2$Ru$_x$Ga. For example, since Mn $4a$ spins are localized, they could play the role of the slow rare earth, while Mn $4c$ have a more delocalized character, and thus play the role of the transition metal. 
In this work we show differently, that Mn$_2$Ru$_x$Ga switches even though its two sublattices show similar demagnetization times, unlocking different demagnetization times as a necessary condition for switching. 
Another hard constrain for switching Mn$_2$Ru$_x$Ga is that it is only possible when the initial temperature lies below $T_M$ \cite{Davies2020,banerjee2019}. Davies et al. \cite{Davies2020} found that this condition is robust and that can be explained assuming that the dynamics is exchange-driven, which has as a consequence that $dM_{4a}/dt = - dM_{4c}/dt$. This condition also holds, when the coupling to the heat-bath is similar for both sublattices and they demagnetize at similar rates. Here, we demonstrate that switching is possible when the initial demagnetization dynamics is dominated by the coupling to the heat-bath instead the exchange relaxation.

\begin{center}
\begin{figure}[t]
\includegraphics[width=1.0\columnwidth]{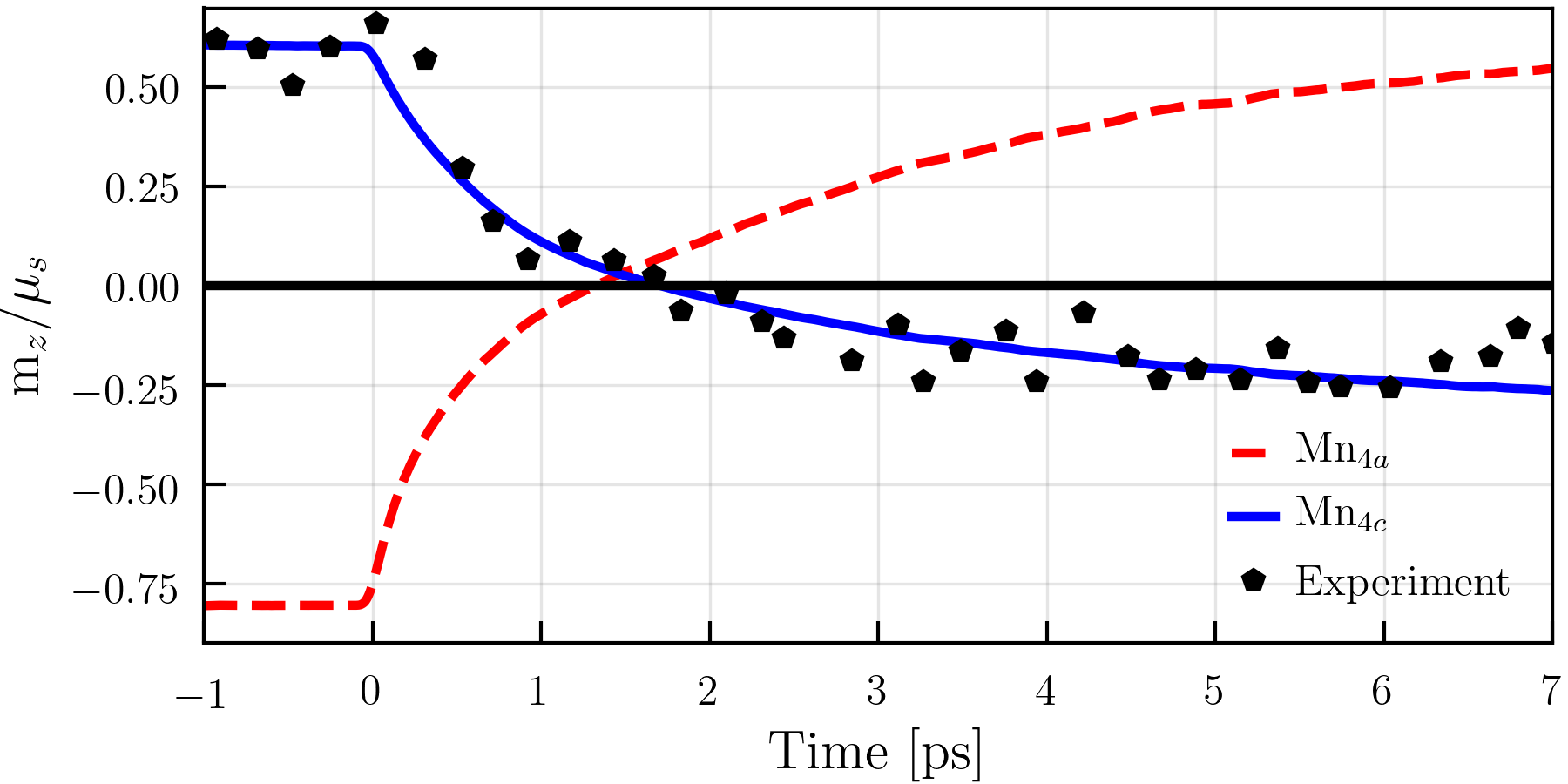}
\caption{Comparison between the experimentally measured Mn$_{4c}$ dynamics for a switching event (points) \cite{Banerjee2021} and our atomistic spin model simulations (lines).}
\label{fig:SwitchingSimulationsExp}
\end{figure}
\end{center}

We use the atomistic stochastic-Landau-Lifshitz-Gilbert equation (sLLG)~\cite{Nowak2007} to describe the magnetization dynamics of the Mn$_2$Ru$_x$Ga alloys and the two-temperature model (TTM) to describe the electron temperature $T_\text{el}$ and the phonon temperature $T_\text{ph}$ (see Supplemental Material (SM) for details)~\cite{Kaganov1957,Chen2006}. 
The parameters defining the TTM for Mn$_2$Ru$_x$Ga are not established yet. 
We use parameters similar to GdFeCo alloys, which are close to those used by Bonfiglio et al. within the 4TM \cite{Bonfiglio2021}. 
The complete set of system parameter used in our model are summarized in table~\ref{table:Parameters} of the SM. 
Within atomistic spin dynamics models, the demagnetization time scales with $\tau \sim \mu_{\rm{at}} /(\alpha T_c)$\cite{Radu2015,MentinkPRL2012}. In two-sublattice magnets, the three parameters, $\alpha,T_c$ and $\mu_{\rm{at}}$ are sublattice-specific. While $T_c$ and $\mu_{\rm{at}}$ are determined by equilibrium properties as discussed before, the value of the damping parameter is related to the coupling to the heat-bath. One method to estimate this value from experiments is to photo-excite magnetization precession. These experiments have been conducted in Mn$_2$Ru$_x$Ga and measured by time resolved Faraday effect as a function of the applied field and temperature. From the decay of the precession the intrinsic damping parameter has been determined to have values smaller than $0.02$ far from $T_{\rm{M}}$ \cite{Bonfiglio2019}.
We chose a damping value ($\alpha_{4c}=0.01$ and $\alpha_{4c}=0.013$) that reproduces the ultrafast magnetization dynamics of the Mn $4c$ sublattice during a switching event as measured by Banerjee et al. \cite{Banerjee2021} and allows switching for $x>0.85$ similar to the experimental findings of Banerjee et al.~\cite{banerjee2019}. 
We note, that Davies et al.~\cite{Davies2020} find switching already for $x=0.75$ which could be reproduced for a different choice of damping values (discussion in the SM).
With this our model is able to quantitatively reproduce the recently measured demagnetization dynamics of the 4$c$ lattice by Banerjee et al. (Fig.~\ref{fig:SwitchingSimulationsExp}).

\begin{center}
\begin{figure}[t]
\includegraphics[width=1\columnwidth]{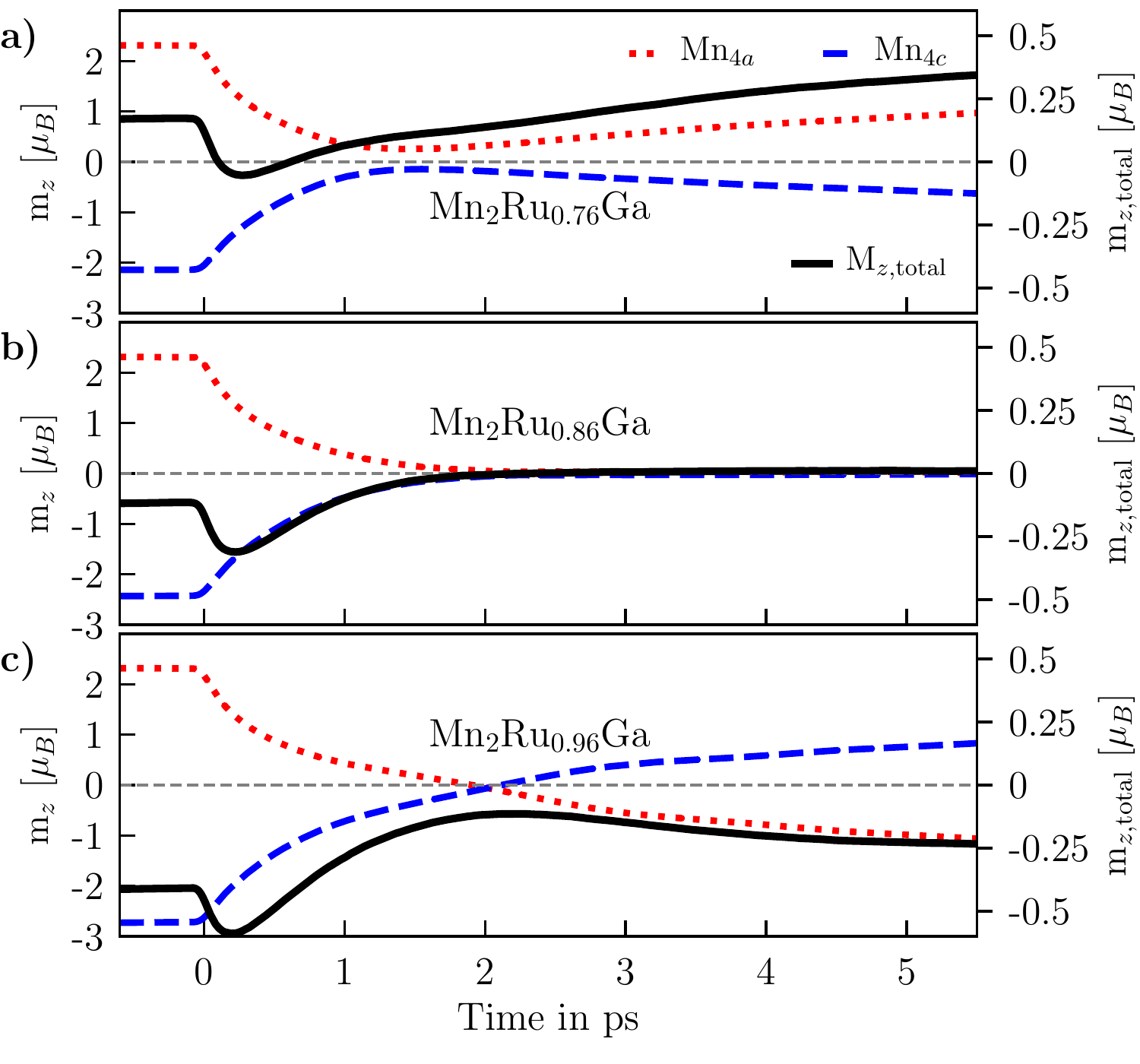}
\caption{Site-specific magnetization dynamics of the $4a$ and $4c$ Mn sites (red, blue; left axis) after a 100 fs laser pulse excitation at $t=0$ (Gaussian pulse peak) for an increasing Ruthenium concentration ($\alpha_{4c} = 0.01$, $\alpha_{4a}=0.013$). The total magnetization ($m_{z,\rm{total}} = m_{z,4a} + m_{z,4c}$) is shown as a black line (right axis). 
All laser parameters were the same for all three simulations. a) shows the no-switching scenario for a Mn$_2$Ru$_{0.76}$Ga alloy.
b) shows the non-deterministic scenario for a Mn$_2$Ru$_{0.86}$Ga alloy, with a prolonged demagnetization state. c) shows the switching scenario for a Mn$_2$Ru$_{0.96}$Ga alloy. 
}
\label{fig:SwitchingSimulations}
\end{figure}
\end{center}
How does our model compare to the current understanding of switching in Mn$_2$Ru$_x$Ga? Previous works~\cite{Davies2020,Bonfiglio2021,Banerjee2021} have suggested, that exchange-driven dynamics dominate the first step of the demagnetization and switching process. This is assumed since the inter-lattice antiferromagnetic exchange is stronger than in (for example) transition metal rare earth alloys. Exchange-relaxation stems from processes driving both sublattices to a mutual equilibrium. Thus, it is unlikely that exchange processes play a role on the first steps of the dynamics. Moreover, since exchange processes describe transfer of angular momentum between sublattices, it is more likely that those processes dominate when the sublattice magnetic order is small rather than when its saturated. Near equilibrium, relaxation by coupling to the heat-bath dominates, while for situations of non-equilibrium and reduced magnetic order, the exchange-relaxation dominates.

Figure~\ref{fig:SwitchingSimulations} shows the site-specific dynamics for three characteristic Ru-concentrations of Mn$_2$Ru$_x$Ga for $x=0.76$ (a), $x=0.86$ (b) and $x=0.96$ (c). These three cases represent alloys below/at/above the threshold of $x=0.9$, which defines the experimentally found switching condition~\cite{banerjee2019}.

A closer look to $m_{z,\rm{total}}$ in Fig. \ref{fig:SwitchingSimulations} indicates that up to the first picosecond $m_{z,\rm{total}}$ is not conserved, although it only changes slightly due to the similar demagnetization dynamics of the Mn $4a$ and $4c$ sublattices (note, that the scale of the right $y$-axis is much smaller than the scale of the left axis). This means that relaxation by coupling to the heat-bath dominates. This relaxation can in turn be interpreted as excitation of ferromagnetic (optical) magnons (following Banerjee and co-workers\cite{Banerjee2021}).
However, as the sublattice magnetization reduces to small values ($\sim$ 1 ps), the value of  $m_{z,\rm{total}}$ stays constant in time.  This means that exchange-relaxation dominates the dynamics (conservation of total angular momentum). This can be interpreted as excitation of antiferromagnetic (acoustic) magnons. Total angular momentum is conserved for a relatively long period of time (few picoseconds).   
Once the system has reached the exchange relaxation dominated regime ($\approx$ 1 ps) the interpretation of switching of Davies and co-workers remains valid. The exchange relaxation results in $|\Delta m_{4a}|=|\Delta m_{4c}|$, but since $|m_{4a,0}| < |m_{4c,0}|$ means that $m_{4a}$ crosses zero before $m_{4c}$ does and therefore giving rise to successful magnetic switching. Differently to the interpretation of Davies et al. \cite{Davies2020} this condition,  $|m_{4a,0}| < |m_{4c,0}|$, holds because the first steps of the sublattice demagnetization are driven by direct coupling to the heat-bath with relaxation at similar speeds.

Our picture can be easily checked. We argue that switching happens if $|m_{4a}| < |m_{4c}|$ when the system enters the exchange relaxation regime, which dominates the dynamics for small values of the sublattice magnetic order. The sign of the total magnetization at that moment can be controlled by the coupling to the heat-bath of each sublattice, namely the value of $\alpha_{4a(4c)}$. 
By increasing $\alpha_{4a}$ the demagnetizaton due to the coupling to the heat-bath increases so that $|m_{4a}| < |m_{4c}|$ can be fullfilled even for smaller values of $x$. The effects of this change of $\alpha_{4a}$ on $m_{\rm{total}}$ would change it from a behaviour as shown in Fig.~\ref{fig:SwitchingSimulations}~a) towards the one in Fig.~\ref{fig:SwitchingSimulations}~c). The full result of this analysis is shown in Fig.~\ref{fig:SwitchingMapSM} in the Supplemental Material. Where we observe that by increasing the value of $\alpha_{4a}$, a wider range of Ru-concentrations switch and thus lowering the switching threshold set by the compensation temperature, ($x=0.90$ in our model).

\begin{center}
\begin{figure}[t]
\includegraphics[width=1.0\columnwidth]{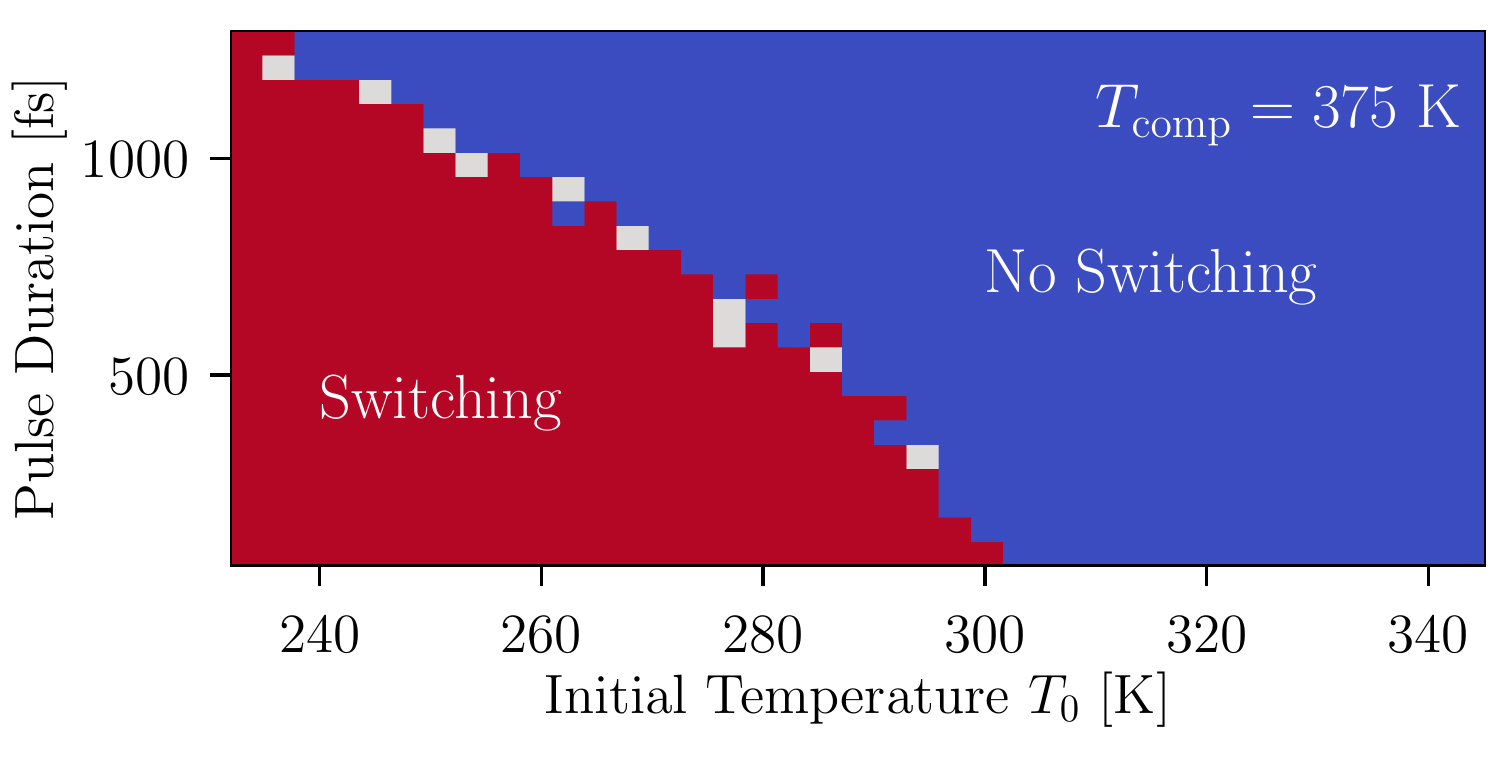}
\caption{Switching behavior as color of Mn$_2$Ru$_{0.98}$Ga as function of the starting temperature $T_0$ and the pulse duration. Red areas indicate switching behavior, blue marks areas without switching and grey indicates a prolonged transient ferromagnetic state or a demagnetized state.}
\label{fig:SwitchingPulseMap}
\end{figure}
\end{center}

Similar to the experimental work of Davies et al.~\cite{Davies2020} we investigate switching as function of the starting temperature $T_0$ and the pulse duration (Fig.~\ref{fig:SwitchingPulseMap}).
Our results compare qualitatively well to the experimental observations of Ref.~\onlinecite{Davies2020}. 
Quantitative comparison between our model, based on the experimental measurements of Banerjee et al.~\cite{banerjee2019}, and Davies et al. are difficult without modifying some of the material parameters used in the model. The main difference between the two experiments is the value of the compensation temperature for similar Ru-concentrations ($T_{M,x=0.75}=370$~K~\cite{Davies2020} and $T_{M,x=0.7}=245$~K~\cite{banerjee2019}).
In order to mimic the conditions of Davies et al., we use Mn$_2$Ru$_{0.98}$Ga, which features a compensation temperature of $T_M=375$ K (experimentally $T_{M,x=0.75}=370$ K). 
The pulse-duration switching threshold decreases as the system base temperature approaches a threshold temperature with a similar slope as experiments. Experimentally this threshold temperature is close to the compensation temperature, while in our model is a bit further. 
However our findings suggest that the condition $T_0 < T_M$ is not a limiting, sufficient condition for switching in Mn$_2$Ru$_x$Ga.

To summarize, we have presented an atomistic spin model able to describe single pulse toggle switching in Mn$_2$Ru$_x$Ga Heusler alloys. 
The parameters of the spin model are based on experimental measurements and can reproduce key material parameters such as $T_c$, $M(T)$ or the Ru-concentration dependence of $T_M$. 
We show that our model is able to quantitatively reproduce measured magnetization dynamics of single pulse toggle switching and we demonstrate that differently to previous understanding toggle switching in Mn$_2$Ru$_x$Ga is possible even when both Mn sublattices demagnetization at very similar rate.

\textbf{Acknowledgments}

We gratefully acknowledge support by the Deutsche Forschungsgemeinschaft through SFB/TRR 227  "Ultrafast Spin Dynamics", Project A08.

\textbf{Data Availability}

The data that support the findings of this study are available from the corresponding author upon reasonable request.

\bibliography{libFJ}

\clearpage 

\widetext 

\section{Supplementary Material}

\subsection{Details of the estimation of the system parameters}
We model Mn$_2$Ru$_x$Ga Heusler alloys based on the Heisenberg spin Hamiltonian: 
 \begin{equation}   
\mathcal{H}= - \sum_{i \neq j} J_{ij} \mathbf{S}_i \cdot \mathbf{S}_j - \sum_{i} d_z S_z^2.
\label{eq:Ham}
\end{equation}
Here $\mathbf{S}_i = \boldsymbol{\mu}_i/\mu_{s,i}$ represents a classical, normalized spin vector at site $i$. Here, $\mu_{s,i}$ is the atomic magnetic moment of the site $i$. 
$\mathbf{S}_i$, couples to its nearest neighboring spin $\mathbf{S}_j$ via the coupling constant $J_{ij}$, and $d_z$ represents the on-site anisotropy with easy-axis along the $z$-axis.

The considered Heisenberg exchange parameters are the ferromagnetic exchange couplings $J_{4a-4a}$ and $J_{4c-4c}$ as well as the antiferromagnetic $J_{4a-4c}$.
The values of the exchange and the uniaxial anisotropy for Mn$_2$Ru$_{0.68}$Ga are based on the values of a molecular field model \cite{Smart1966} derived in Ref.~\onlinecite{Fowley2018}.
The exchange parameters are adjusted to yield a Curie temperature of about 630~K  which is at the upper end of reported Curie temperatures, for instance, it is close to the value reported in Ref.~\onlinecite{Fowley2018} of $T_c = 625$~K for $x=0.68$, where $x$ stands for Ru-concentration.
Additional experiments find an approximately constant Curie temperature of 450 K for $x>0.5$ ~\cite{Kurt2014}, whereas Ref.~\onlinecite{Bonfiglio2021} reports $T_c = 550 $ K for $x=0.7$. 
Furthermore the T$_c$ measured in Ref.~\onlinecite{Kurt2014} features a peak around $x=0.4$ which could explain why our parameters only reproduce reproduce $T_{\text{comp}}$ for $x > 0.68$. 
Thus, a dependence of the exchange parameters on Ru-concentration may therefore be possible for lower values of $x$. 
Since we are mostly interested in Ru-concentration for which switching has been experimentally demonstrated ($x>0.7$), we restrict our study to Ru concentrations of $x=0.6-1.0$. 
We note that we have increased $J_{4a-4a}$ by 10\% to reproduce the magnetization compensation temperature as reported in Ref.~\cite{banerjee2019} while keeping T$_c = 630$ K. Using exchange constants with the ratios as stated in Ref.~\onlinecite{Fowley2018} couldn't reproduce the experimentally measured Mn$_{4c}$ dynamics of C. Banerjee et al.~\cite{Banerjee2021} that are shown in Fig.~\ref{fig:SwitchingSimulationsExp} (main text). However, the fact that there are  reports of largely varying different Curie temperatures \cite{Fowley2018,Kurt2014,Bonfiglio2021} and compensation temperature~\cite{banerjee2019,Davies2020} indicate that there seems to be a wide spectrum of valid exchange constants between different experiments, so that an adjustment of 10\% of one of the parameters seems reasonable.

The uniaxial anisotropy is also considered to be site-specific, in particular, the anisotropy energy density K$_{z,4c} = 216$ kJm$^{-3}$ and K$_{z,4a} = 0$ kJm$^{-3}$ taken from Ref.~\cite{Fowley2018}. Assuming a unit cell size of ($\approx 0.6$ nm)$^3$ (\cite{Thiyagarajah2015}) we obtain $d_{z,4c} = 1.1664 \cdot 10^{-23}$ J as on-site anisotropy and $d_{z,4a} = 0$.
The anisotropy is included to yield an alignment along the $z$-axis after demagnetization or switching. Since it is much smaller the Heisenberg exchange it does not have meaningful impact on the switching itself.
Furthermore ref.~\onlinecite{Bonfiglio2019} finds, based on XMCD experiments, different g-factors, $g_{4a} = 2.05$ and $g_{4c} = 2.00$ for the 4$a$- and 4$c$-sublattice. However since both $\gamma$ and $\mu_s$ are proportional to $g_i$ this does not enter the LLG.

The spin dynamics of this system are described by the atomistic stochastic-Landau-Lifshitz-Gilbert equation (sLLG)~\cite{Nowak2007}
\begin{equation}
\frac{(1+\alpha_i^2)\mu_{s,i}}{\gamma}\frac{\partial \mathbf{S}_i}{\partial t} = - \left( \mathbf{S}_i \times \mathbf{H}_i \right) - \alpha_i  \left( \mathbf{S}_i \times \left(\mathbf{S}_i \times \mathbf{H}_i \right) \right).
\label{eq:llg}
\end{equation}
Where $\gamma$ represents the gyromagnetic ratio and $\alpha_i$ is a site-specific atomic damping parameter. 
Here, we also draw on experimental observations to estimate the values of the damping parameters. Photo-excited spin precession was observed by time resolved Faraday effect as a function of the applied field and temperature. From the decay of the precession the intrinsic damping parameter was also determined to have values smaller than $\alpha=0.02$, far from compensation \cite{Bonfiglio2019}. Here, we decided to use $\alpha_{4c}=0.01$ and $\alpha_{4c}=0.013$ since it reproduces experimental observations as discussed in the main text.
The temperature dynamics are described using the two-temperature model (TTM) that describes the electron temperature $T_\text{el}$ and the phonon temperature $T_\text{ph}$ via a pair of two coupled differential equations~\cite{Kaganov1957,Chen2006}:
\begin{align}
C_{\rm{el}} \frac{\partial T_{\rm{el}}}{\partial t} &= -g_{\rm{ep}}\left( T_{\rm{el}} - T_{\rm{ph}} \right) + P_{l}(t) \\
C_{\rm{ph}} \frac{\partial T_{\rm{ph}}}{\partial t} &= +g_{\rm{ep}}\left( T_{\rm{el}} - T_{\rm{ph}} \right).
\label{eq:2TM}
\end{align}
$C_\text{el}$ and $C_\text{ph}$ represent the specific heat of the electron- and phonon system and $P_l(t)$ describes the absorbed energy of the electron system, coming from the laser. 
Since experimental data on the electron and phonon temperature dynamics is missing, we use similar parameters to typical GdFeCo values for our TTM and similar ones to the experimental work by Bonfiglio et al. (Ref.~\cite{Bonfiglio2021}).
Table~\ref{table:TTM-Parameters} provides an overview of the used TTM-parameters in comparison to Ref.~\cite{Bonfiglio2021} and to parameters used in GdFeCo from Ref.~\onlinecite{Barker2013} and Ref.~\onlinecite{Mekonnen2013}.
\begin{table}[h!]
\caption{Two temperature model parameters comparison between GdFeCo and Mn$_2$Ru$_x$Ga between different sources.}
\label{table:TTM-Parameters}
\begin{tabular}{|ll||c|c||c|c|}
\hline 
TTM & Unit & Mn$_2$Ru$_x$Ga & Mn$_2$Ru$_x$Ga\cite{Bonfiglio2021} & GdFeCo\cite{Mekonnen2013} & GdFeCo\cite{Barker2013} \cr
\hline
C$_\text{ph}$ & J/m$^3$K & 3 $\times 10^6$ & 2.27 $\times 10^6$ & 3 $\times 10^6$ & 3 $\times 10^6$
\\
g$_\text{ph}$ & J/m$^3$Ks & 6 $\times 10^{17}$ & 8 $\times 10^{17}$ &  2 $\times 10^{17}$ &  17 $\times 10^{17}$\\
$\gamma$ & J/m$^3$K$^2$ & 350 & 484  & 714 & 700\\\hline
\end{tabular}
\end{table}

The laser pulse  is assumed to be Gaussian shaped with a FWHM of 100 fs.
The electron temperature $T_{\rm{el}}$ yielding from the TTM is used to scale the temperature effects in the spin system. This is done by including a Langevin thermostat, which adds an effective field-like stochastic term $ \boldsymbol{\zeta}_i$ to the effective field $\mathbf{H}_i= \boldsymbol{\zeta}_i(t) - \frac{\partial \mathcal{H}}{\partial \mathbf{S}_i}$ with white noise properties~\cite{Atxitia2009}:
\begin{equation}
\langle \boldsymbol{\zeta}_i(t) \rangle = 0 \quad \text{and} \quad \langle \boldsymbol{\zeta}_i(0) \boldsymbol{\zeta}_j(t) \rangle = 2 \alpha_i k_\text{B} T_{\rm{el}} \mu_{s,i} \delta_{ij}\delta(t)/\gamma.
\label{eq:noise-correlator}
\end{equation}

The complete set of system parameter used in our model are summarized in table~\ref{table:Parameters}. 

\begin{table}[h!]
\caption{Table of the Heisenberg spin Hamiltonian parameters (left) and the two temperature model (TTM) (right).}
\label{table:Parameters}
\begin{tabular}{ l |lllll||l|ll}
\hline 
\hline
$\mathcal{H}$& &&Value&& Unit &TTM & &Unit\\ \hline
$J_{4a-4a}$ & &&$1.28 $&$\times 10^{-21}$ &[J]& C$_{\rm{ph}}$& $3\times 10^{6}$&[J/Km$^3]$\\ 
$J_{4c-4c}$ & &&$4.0 $&$\times 10^{-22}$ &[J]& C$_{\rm{el}}$ &$\gamma_\text{el} \cdot T_{\rm{e}}$&[J/Km$^3]$\\ 
$J_{4a-4c}$ & &$-$&$4.85 $&$\times 10^{-22}$ &[J]& $\gamma_\text{el}$& 350& [J/K$^2$m$^3$]\\  
$\gamma$ & &&$1.76$&$\times 10^{-21}$&$[\frac{1}{\text{Ts}}]$&g$_{\rm{ep}}$&$6\times 10^{17}$&[J/sKm$^3]$\\
$d_z$ & &&$1.17$&$\times 10^{-23}$&[J]&\\
$\mu_{\rm{s,4a}}$& && 2.88 &&$[\mu_{\rm{B}}]$&&\\
$\mu_{\rm{s,4c}}$& && $4.71\cdot x$&&$[\mu_{\rm{B}}]$&&\\ 
$\alpha_{4a}$& & &$0.013$&&&&\\
$\alpha_{4c}$& & &$0.01$&&&&\\ \hline \hline
\end{tabular}
\end{table}

\subsection{Switching behaviour in dependence of Gilbert damping parameters}

\begin{center}
\begin{figure}[t]
\includegraphics[width=0.85\columnwidth]{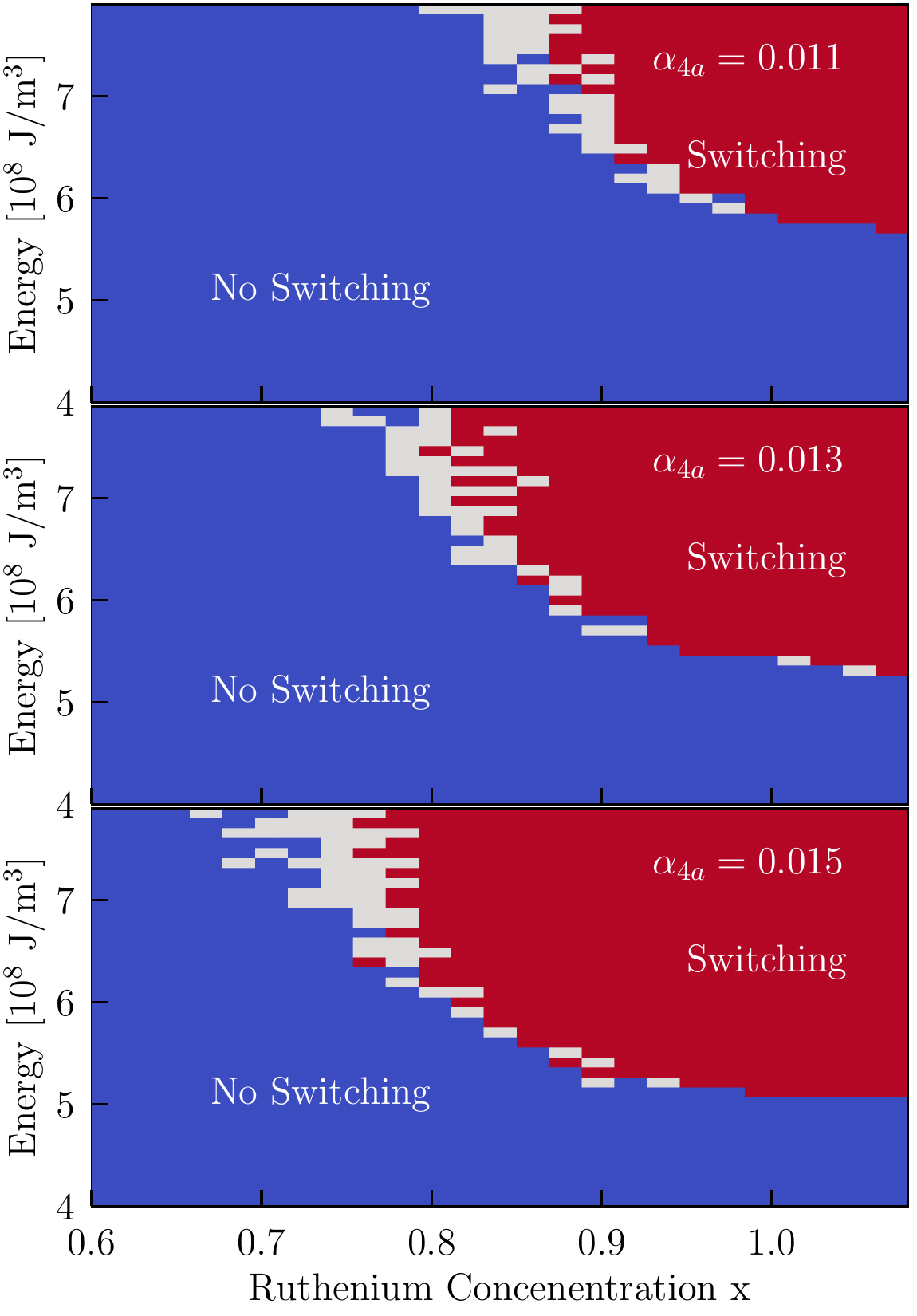}
\caption{Switching behavior as color of Mn$_2$Ru$_x$Ga alloys as function of the Ruthenium concentration $x$ and the absorbed laser energy for different dampings $\alpha_{4a}$ of the $4a$-sublattice. Red areas indicate switching behavior, blue marks areas without switching and grey indicates a prolonged transient ferromagnetic state or a demagnetized state. The damping $\alpha_{4c}=0.01$ of the $4c$-sublattice was kept constant while $\alpha_{4a}$ was varied from $\alpha_{4a}=0.011$ (top) to $\alpha_{4a}=0.013$ (middle) to $\alpha_{4a}=0.015$ (bottom).}
\label{fig:SwitchingMapSM}
\end{figure}
\end{center}

Figure~\ref{fig:SwitchingMapSM} shows the switching behavior of Mn$_2$Ru$_x$Ga alloys as function of the Ru-concentration $x$ and the absorbed laser energy for different dampings $\alpha_{4a}$. Red areas indicate switching behavior, blue marks non-switched simulations and grey areas indicate simulations with a prolonged transient ferromagnetic state or a demagnetized state.
The simulation was counted as switched, when the $4a$-sublattice crossed $m_z=0$ and reached a threshold of $m_{z,4a} < -0.12$ after 15 ps (starting at positive $m_z$ values). Otherwise it was counted as demagnetized if $|m_{z,4a}| < 0.12$, or remagnetized if $m_{z,4a} >= 0.12$ (see Fig.~\ref{fig:SwitchingSimulations} for examples).

The damping $\alpha_{4c}=0.01$  was kept constant while $\alpha_{4a}$ was varied from $\alpha_{4a}=0.011$ (top) to $\alpha_{4a}=0.013$ (middle) to $\alpha_{4a}=0.015$ (bottom).
Figure~\ref{fig:SwitchingMapSM}, shows clearly distinguish behaviors for the simulated alloys depending on the Ru-concentration $x$.
For low absorbed laser energies below $5.5 \cdot 10^8$ J/m$^3$ no switching occurs for all Ru concentrations considered here. This is due to the insufficient energy to temporarily demagnetize both sublattices.
For all three cases, we find that for low Ruthenium concentrations below a damping-dependent threshold value the alloy does not switch, independent of the laser energy. 
Only above that threshold we find deterministic switching.
In the top panel with $\alpha_{4c}=0.01$ and $\alpha_{4a}=0.011$, we find the threshold Ru-concentration for switching to be around $x\approx 0.9-0.95$. When the damping of the $4a$ sublattice is increased to $\alpha_{4a}=0.013$ the threshold moves to $x\approx0.85-0.9$, which approximately corresponds to the switching threshold found in Ref.~\onlinecite{banerjee2019}. Finally, for $\alpha_{4a}=0.015$ (bottom) the switching threshold decreases to $x\approx0.8$.
Therefore we find, that by increasing the element specific damping discrepancy the switching threshold moves towards lower Ruthenium concentrations. Our results compare best to the experiments by Banerjee and co-workers (threshold around $x=0.8-0.9$) \cite{banerjee2019} when choosing $\alpha_{4c}=0.01$ and $\alpha_{4a}=0.013$ (Fig. \ref{fig:SwitchingMapSM} middle).
We note that in our model the only parameter that directly depends on $x$ is $\mu_{s,4c}$, which impacts the speed of the $4c$-sublattice. 
Our model shows that by increasing $x$,  $\mu_{s,4c}$ also increases and in turn the sublattice demagnetization speed continuously slows down, up to the point where the demagnetization speed difference in the $4a$- and $4c$-sublattice is large enough to enable switching behavior. The model also shows that this relatively different demagnetization speed can also be controlled by the intrinsic site-dependent damping parameters, which  influences the $x$-dependent threshold between switching and non switching behavior.

\end{document}